\documentclass[lettersize,journal]{IEEEtran}
\usepackage{amsmath}
\usepackage{amsmath,amsfonts}
\usepackage{algorithmic}
\usepackage{algorithm}
\usepackage{empheq}
\usepackage{array}
\usepackage[caption=false,font=footnotesize,labelfont=rm,textfont=rm]{subfig}
\usepackage{textcomp}
\usepackage{stfloats}
\usepackage{url}
\usepackage{verbatim}
\usepackage{graphicx}
\usepackage{cite}
\usepackage[T1]{fontenc}
\usepackage{booktabs,multirow}
\usepackage{pifont}
\usepackage{CJKutf8}
\usepackage{hyperref} 
\hypersetup{hidelinks,
  colorlinks=ture,
  linkcolor=black
  }

\begin{document}
\begin{CJK}{UTF8}{gbsn}
\title{Joint ADS-B in 5G for Hierarchical Aerial Networks: Performance Analysis and Optimization}

\author{Ziye Jia\textsuperscript{1,2}, Yiyang Liao\textsuperscript{1}, Chao Dong\textsuperscript{1}, Lijun He\textsuperscript{3},   
Qihui Wu\textsuperscript{1}, and Lei Zhang\textsuperscript{1}\\
\textsuperscript{1}The Key Laboratory of Dynamic Cognitive System of Electromagnetic Spectrum Space, Ministry of Industry \\
and Information Technology, Nanjing University of Aeronautics and Astronautics, Nanjing, Jiangsu, 210016 \\
\textsuperscript{2}National Mobile Communications Research Laboratory, Southeast University, Nanjing, Jiangsu, 211111\\
\textsuperscript{3}School of software, Northwestern Polytechnical University, Xi'an, Shanxi, 710129\\
\{jiaziye, liaoyiyang, dch, wuqihui, Zhang\_lei\}@nuaa.edu.cn, lijunhe@nwpu.edu.cn
}
\maketitle
\pagestyle{empty}  
\thispagestyle{empty} 
\begin{abstract}
        Unmanned aerial vehicles (UAVs) are widely applied in multiple fields, which emphasizes
        the challenge of obtaining UAV flight information to ensure the airspace safety. UAVs equipped with automatic dependent surveillance-broadcast (ADS-B) devices are capable of sending flight information to nearby aircrafts and ground stations (GSs). However, the saturation of limited frequency bands of ADS-B leads to interferences among UAVs and impairs the monitoring performance of GS to civil planes. To address this issue, the integration of the 5th generation mobile communication technology (5G) with ADS-B is proposed for UAV operations in this paper. Specifically, a hierarchical structure is proposed, in which the high-altitude central UAV is equipped with ADS-B and the low-altitude central UAV utilizes 5G modules to transmit flight information. Meanwhile, based on the mobile edge computing technique, the flight information of sub-UAVs is offloaded to the central UAV for further processing, and then transmitted to GS. We present the deterministic model and stochastic geometry based model to build the air-to-ground channel and air-to-air channel, respectively. The effectiveness of the proposed monitoring system is verified via simulations and experiments. This research contributes to improving the airspace safety and advancing the air traffic flow management.
\end{abstract}

\vspace{-0.4cm} 
\section{Introduction}
\IEEEPARstart{A}{s} the aviation technology advances, unmanned aerial vehicles (UAVs) become increasingly prevalent. Due to the characteristics of the dependability, versatility, and adaptability, UAVs are widely used in areas such as the agriculture, aerial photography, logistics, surveillance, emergency response, and environmental conservation\cite{ref1}. Nevertheless, the increment of UAVs also brings significant concerns regarding the flight safety. It is imperative to enhance the flight control and safety monitoring to ensure that UAV operations do not disturb the public safety or individual privacy\cite{ref2}. In response to these challenges, this paper proposes the hierarchical management structure for UAV networks, with the cooperation of automatic dependent surveillance-broadcast (ADS-B) \cite{ref3} systems into the 5th generation mobile communication technologies (5G)\cite{ref4}, \cite{ref5}.

\par In particular, the ADS-B system is composed of multiple ground stations (GSs) and airborne stations, which works at a specific frequency band, i.e., 1090MHz\cite{ref6}. An aircraft equipped with ADS-B can automatically broadcast its position information to nearby aircrafts and GSs \cite{ref7}. However, due to the limited frequency band, excessive UAVs equipped with ADS-B interfere in the surveillance of GS towards civil planes. The typical impact is intensifying the collision of ADS-B packets, which leads to packets loss and impairs the monitoring performance of GS towards civil planes\cite{ref8}. Further, the 5G technology, characterized by ultra-high speed, minimal latency, and extensive bandwidth, can satisfy the wireless communication needs for the UAV flight control and operational tasks\cite{ref9}. The integration of 5G modules with UAVs facilitates communication abilities between UAVs and GS\cite{ref10}, which assists GS in obtaining up-to-date flight information promptly. However, using 5G modules alone is difficult to integrate the monitoring system with the mature civil aviation control system. Hence, in order to alleviate the impacts on monitoring performance, ensure timely acquisition of the flight information, and expand the monitoring capacity of GS, we consider the cooperation of ADS-B with 5G. Furthermore, the mobile edge computing (MEC) techniques are deployed on a couple of UAVs, which progress the comprehensive performance of the proposed network.

\par There exist some works about UAV equipped with ADS-B or 5G. For example, \cite{ref6} proposes a system for UAV surveillance based on ADS-B and leverage recurrent long short-term memory to minimize the prediction error of UAV trajectory. \cite{ref8} analyses the impact of UAVs equipped with ADS-B on the civil planes at the frequency of 1090MHz, verifying the access capacity of the network layer.
\cite{ref11} points out that 5G fully meets the requirements for the wireless communication performance of UAVs flight control and missions, achieving real-time interaction between UAVs and GS. \cite{ref12} develops a novel trust-based security scheme for 5G UAV communication systems, improving the communication performance and evaluating the reliability. Besides, there exist some related works about the deployment of MEC on UAVs, to advance the performance of UAV communications by appropriately utilizing computing resources at the edge of networks\cite{ref13}. For example, \cite{ref14} envisions a pre-dispatch UAV-assisted vehicular edge computing networks system, and the UAVs play roles as temporary base stations to cope with the demand of vehicles in multiple traffic jams. \cite{ref15} proposes a joint communication and computation optimization network model of UAV swarms, which leverages MEC to decrease the response delay and increase the efficiency of network resources utilization. As far as the authors' knowledge, the researches solve the problem of how to supervise UAVs, but they seldom consider the cooperation system with ADS-B in 5G, which can increase the monitoring capacity, advance the performance of surveillance, and help the integrated management for air traffic control.

\par In this work, we propose a hierarchical aerial management structure for UAV networks, with the cooperation of ADS-B and 5G. The hierarchical system includes two types of communication channels, i.e., the air-to-air (A2A) channel and air-to-ground (A2G) channel. Besides, the central UAVs are facilitated with MEC ability, to increase the monitoring performance of the system. The flight information of sub-UAVs is initially relayed to the central UAVs, and then transmitted to GS after MEC based processing. Such integration facilitates the transmission of the flight information from UAVs to GS via ADS-B or 5G modules, enabling effective cooperative aerial surveillance.

\par The rest of this paper is organized as follows. Section \ref{S2} introduces the system model. The stochastic performance analysis is presented in Section \ref{S3}. Section \ref{S4} describes the mechanisms of on-board processing based on MEC. Section \ref{S5} provides the simulation results and corresponding analyses.  Finally, Section \ref{S6} draws the conclusions.
\vspace{-0.2cm}

\section{System Model}\label{S2}

\begin{figure}[t]
        \centerline{\includegraphics[width=1\linewidth]{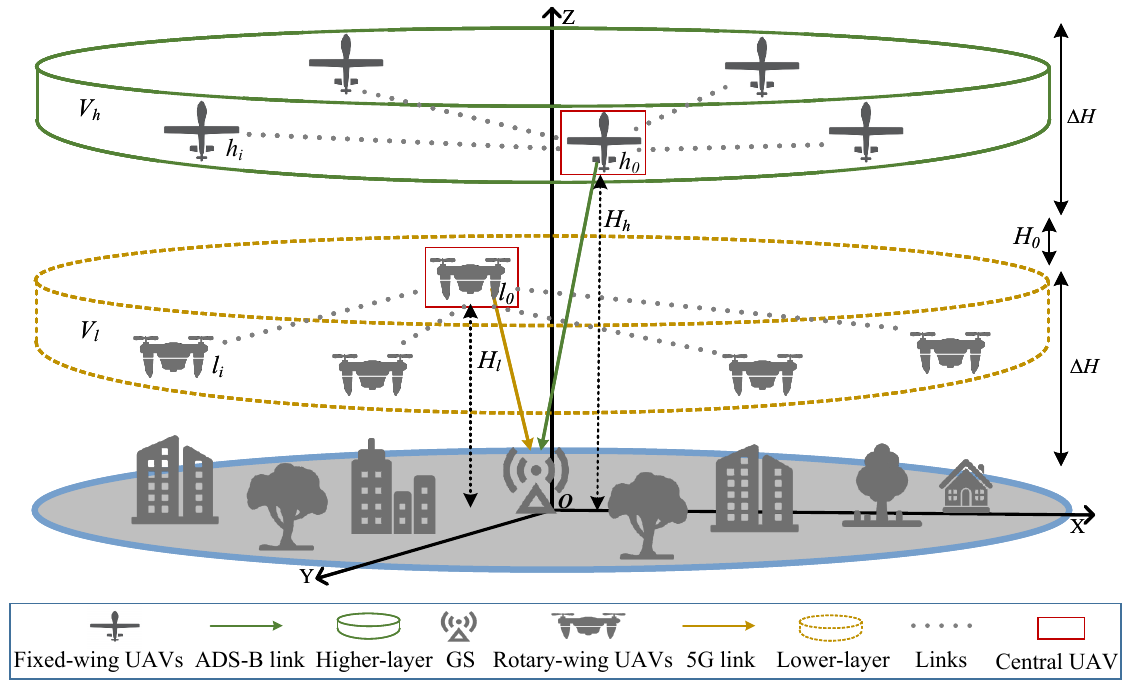}}
        \caption{Hierarchical aerial network model.}
        \label{f1}
\end{figure}

 Fig. \ref{f1} depicts the hierarchical aerial network model. In detail, the rotary-wing UAVs with density of $\lambda_l$ are randomly distributed in the low-altitude airspace $V_l$, corresponding to the orange dotted line area. Due to the better flying endurance, the fixed-wing UAVs with density of $\lambda_h$ scatter in high-altitude airspace $V_h$, shown as the green solid line area. Besides, a central UAV is deployed in each layer. In detail, the low-altitude central UAV is equipped with 5G modules, while the high-altitude central UAV transmits the flight information through the ADS-B system. $H_h$ and $H_l$ are regarded as the altitude of the central UAV $h_0$ and central UAV $l_0$, respectively. Moreover, $\Delta H$ is the height of both layers. Furthermore, there is an isolation layer with height $H_0$ between the two airspaces, aiming to reduce the interference among the UAVs from different airspaces. Besides, it is assumed that there is a GS located at \textsl{\textbf{O}}(0, 0, 0) in the space. GS can receive the flight information from both the ADS-B and 5G modules. Besides, to improve the monitoring performance, sub-UAVs do not establish direct contacts with GS.
 
 \vspace{-0.2cm}  
 \subsection{UAV Model}
\par We denote the set of high-altitude UAVs as $\mathcal H=\{h_0,h_1,...,h_i,...,h_u\}$ and $i\in (1,u)$. Moreover, we indicate the set of low-altitude UAVs as $\mathcal L=\{l_0,l_1,...,l_j,...,l_v\}$ and $j\in (1,v)$. $u$ and $v$ represent the number of the corresponding sub-UAVs. In detail, $h_0$ in $\mathcal H$ symbolizes the central UAV in the high-altitude airspace while $l_0$ in $\mathcal L$ indicates the central UAV in the low-altitude airspace. Additionally, the coordinate of the $i$-$th$ UAV in set $\mathcal H$ is $(x_{h_i}$, $y_{h_i}$, $z_{h_i})$ and the coordinate of the $j$-$th$ UAV in set $\mathcal L$ is $(x_{l_j}$, $y_{l_j}$, $z_{l_j})$. The euclidean distance between UAV $h_i$ and $h_0$ is $d_{h_i}$=$\sqrt{(x_{h_i}-x_{h_0})^2+(y_{h_i}-y_{h_0})^2+(z_{h_i}-z_{h_0})^2}$, and the euclidean distance between UAV $l_j$ and $l_0$ is $d_{l_j}$=$\sqrt{(x_{l_j}-x_{l_0})^2+(y_{l_j}-y_{l_0})^2+(z_{l_j}-z_{l_0})^2}$.  $P_s$ represents the transmitting power of all sub-UAVs. In addition, $P_c$ is the transmitting power of the central UAV, which is further divided to $P_{ch}$ of the central UAV $h_0$ and $P_{cl}$ of the central UAV $l_0$, respectively. $G_g$ and $G_a$ represent the total gain of the A2G channel and A2A channel, respectively. Besides, $P_s$, $G_g$, and $G_a$ are same for all links in the model. 
\vspace{-0.3cm}  

\subsection{A2G Channel Model} 

\begin{figure}[t]
        \centerline{\includegraphics[width=0.85\linewidth]{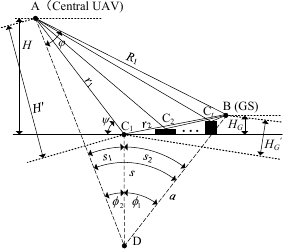}}
        \caption{A2G channel.}
        \label{f2}
\end{figure}

Since the central UAV acts as the aerial station and hovers within a confined area, we employ a deterministic model to establish A2G channels, which has the merits of strong pertinence, precise fit and high accuracy. In the case of long distances, the shape of earth is approximately equivalent to a sphere of radius $a$\cite{ref16}. Therefore, we model the path loss into a curved-earth multi-rays (CEMR) model, as shown in Fig. \ref{f2}. Moreover, the black rectangles represent the reflectors. The distance of LoS component between A and B is $R_1$. Additionally, we take the non-line-of-sight (NLoS) component passing ${\rm C_1}$ for example, whose distance is $R_2$=$r_1$+$r_2$. Furthermore, D is regarded as the center of earth and ${\rm C}_t$ is supposed to be the $t$-$th$ reflection point. Initially, we set the height of the central UAV as $H$, and assume the distance between the central UAV and the tangential plane at ${\rm C_1}$ as $H'$. Specifically, when concerning the high-altitude airspace, $H$ refers to $H_h$. Besides, $H$ conducts as $H_l$ when involving the low-altitude airspace. Subsequently, $H_G$ expresses the height of GS. $H_G'$ declares the distance between GS and the tangential plane at ${\rm C_1}$. Since we arrange the central UAVs to hover in a small area above the GS, the horizontal distance between them is ignored in relation to the vertical altitude, i.e., $H\approx H'$ and $H_G\approx H_G'$. In triangle ABD, according to the law of cosine, we have
\begin{equation} 
        R_1^2=(a+H')^2+(a+H_G')^2-2(a+H')(a+H_G')\cos(\phi),
        \label{eq1}
\end{equation}
where $\phi$=$\phi_1$+$\phi_2$. Besides, $s$=$a\phi$=$s_1$+$s_2$.  To facilitate the calculation, we introduce three intermediate variables $\omega_1$, $\omega_2$, and $\omega_3$:
\begin{equation}
        \begin{cases} 
        \omega_1=\frac{s^2}{4a(H'+H_G')} \\ 
        \omega_2=\frac{H'-H_G'}{H'+H_G'} \\
        \omega_3=2\sqrt{\frac{\omega_1+1}{3\omega_1}}\cos\bigg\{\frac{\pi}{3}+\frac{1}{3}\arccos\bigg[\frac{3\omega_2}{2}\sqrt{\frac{3\omega_1}{(\omega_1+1)^3}}\bigg]\bigg\}. 
        \label{eq2}
        \end{cases} 
\end{equation}
By combining (\ref{eq1}), (\ref{eq2}) and $s$=$a\phi$, we obtain
\begin{equation}
        \begin{cases}
        s_1=\frac{s(1+\omega_3)}{2}, \\ 
        s_2=s-s_1, \\
        \phi_1=\frac{s_1}{a}.
        \end{cases} 
\end{equation}
The phase difference of the signal between path AB and path A${\rm C_1}$B is $\Delta \varphi_1$, calculated as 
\begin{equation}
        \Delta \varphi_1=\frac{2\pi\Delta s}{\lambda}, 
\end{equation}
where $\Delta s$=$2s_1s_2\psi^2/s$, $\psi$=$(H'+H_G')[1-\omega_1(1+\omega_2^2)]/s$ and $\lambda$ is the wavelength. In detail, $\lambda$=$c/f$, in which $c$ is the velocity of light and $f$ is the frequency. Additionally, we assume that the reflection coefficient of earth is $\Gamma_\perp $. Since we consider all UAVs are equipped with vertically polarized antennas, $\Gamma_\perp$ is simplified as
\begin{equation} 
        \Gamma_\perp=\frac{(\varepsilon_r-jb)\sin\psi-\sqrt{(\varepsilon_r-jb)-\cos\psi} }{(\varepsilon_r-jb)\sin\psi+\sqrt{(\varepsilon_r-jb)-\cos\psi}}, 
\end{equation}
where $\varepsilon_r$ is the relative dielectric constant and $j^2$=$-1$. Besides, $b$=$\sigma/(2\pi f\varepsilon_0)$, where $\sigma$ is the electric conductivity, and $\varepsilon_0$ is the dielectric constant.
Besides, the reflection coefficient of point ${\rm C_1}$ is $\Gamma_{1}$=$|\Gamma_{1}|e^{j\phi_{1}}$=$\mathcal{D} \Gamma_\perp$, where $\mathcal{D}$ is the divergence factor. Further, $\mathcal{D}$ is defined as
\begin{equation}
  \mathcal{D}=\bigg[1+\frac{2r_1r_2}{a(r_1+r_2)\sin\psi }\bigg]^{-\frac{1}{2}},
\end{equation}
where $r_1$ is obtained from the law of cosine in triangle A${\rm C_1}$D, i.e., ${r_1}^2$=$(a$+$H')^2$+$a^2$-$2a(a$+$H')\cos\phi_2$ and $r_2$ is obtained from the law of cosine in triangle ${\rm C_1}$BD, i.e., ${r_2}^2$=$(a$+$H_G')^2$+$a^2$-$2a(a$+$H_G')\cos\phi_1$. $E_{LoS}$ and $E_{NLoSt}$ respectively symbolize the signal strength of the LoS path and the $t$-$th$ NLoS path, and $|E_{LoS}|^2$=$P_cG_t$, where $G_t$ denotes the transmitter gain of the central UAV. Hence, the signal strength $E_G$ received by GS is the vector sum of the signals at point B, whose analytic form is 
\begin{align}
        E_G&=E_{LoS}+\sum_{t = 1}^{\lfloor\frac{\pi}{2\varrho }\rfloor}  E_{NLoSt} \nonumber\\ 
        &=E_{LoS}\bigg[1+\sum_{t = 1}^{\lfloor\frac{\pi}{2\varrho }\rfloor}  |\Gamma_t|e^{-j(\Delta \varphi_t-\phi_t)}\bigg]\nonumber\\ 
        &=E_{LoS}\bigg\{\bigg[1+\sum_{t = 1}^{\lfloor\frac{\pi}{2\varrho }\rfloor}|\Gamma_t|\cos(\Delta \varphi_t-\phi_t)\bigg]\nonumber\\
        &-j\sum_{t = 1}^{\lfloor\frac{\pi}{2\varrho }\rfloor}|\Gamma_t|\sin(\Delta \varphi_t-\phi_t)\bigg\},
\end{align}
where $\varrho$ denotes the beamwidth.
Moreover, the power $P_G$ of the signal received by GS is formulated as
\begin{equation}
        P_G=\frac{\|E_G\|^2G_r\lambda^2}{(4\pi R_1)^2}, 
\end{equation}
where $G_r$ is the receiver gain at GS and $G_tG_r$=$G_g$. Hence, the path loss of the A2G communication system is defined as
\begin{equation}
        PL=-10{\rm log}_{10}\bigg(\frac{P_G}{P_c}\bigg),
        \label{eq10}
\end{equation}
which can be further subdivided into $PL_l$=$-10{\rm log}(P_G/P_{cl})$, and $PL_h$=$-10{\rm log}(P_G/P_{ch})$.   
\vspace{-0.4cm}  
\subsection{A2A Channel Model}
Since the sub-UAVs perform mobility and maneuverability during the flight, we establish the connection from them to the central UAV by the stochastic channel based on the stochastic geometry (SG)\cite{ref17}, which has the advantages of universality and flexibility. Considering the hierarchical airspace, we set up an isolation layer between the airspaces, and only the interference in the same airspace is considered for the interferences among the sub-UAVs. Considering the high-altitude airspace, the path loss from the $i$-$th$ sub-UAV to the central UAV is proportional to $d_{h_i}^{-\delta}$, where $d_{h_i}$ denotes the distance between the sub-UAVs and central UAV. $\delta$ indicates the path loss index. Besides, $\rho_{h_i}$ follows an exponential distribution with mean value of 1, indicating the gain of small scale fading channel. Similarly, Gaussian white noise $N$ is added to the model, i.e., $N$=$n_0B$, where $n_0$ is the noise power density and $B$ is the system bandwidth. We leverage $\gamma$ to represent SINR. $\gamma_{h_i}$, the desired signal sent by the $i$-$th$ sub-UAV $h_i$ in set $\mathcal H$ to the central UAV $h_0$, is calculated as

\begin{equation} 
  \gamma_{h_i}=\frac{{P_s}{G_a}{\rho_{h_i}}{d_{h_i}^{-\delta}}}{N+{P_s}S_{\mathcal {H}\backslash \{{h_i}\}}},
\end{equation}

\noindent where

\begin{equation} 
S_{\mathcal{H}\backslash \{{h_i}\}}=\sum_{{h}\in \mathcal{H}\backslash \{{h_i}\}}{G_a}{\rho_{h_i}}{d_{h_i}^{-\delta}}.
\end{equation}
\noindent Similarly, as for the low-altitude airspace, SINR $\gamma_{l_j}$ of the desired signal sent by the $j$-$th$ sub-UAV $l_j$ in set $\mathcal L$ to the central UAV $l_0$ is  
\begin{equation} 
        \gamma_{l_j}=\frac{{P_s}{G_a}{\rho_{l_j}}{d_{l_j}^{-\delta}}}{N+{P_s}S_{\mathcal {L}\backslash \{{l_j}\}}},
      \end{equation}
      
\noindent in which 
\begin{equation} 
      S_{\mathcal{L}\backslash \{{l_j}\}}=\sum_{{l}\in \mathcal{L}\backslash \{{l_j}\}}{G_a}{\rho_{l_j}}{d_{l_j}^{-\delta}}.
\end{equation}

\section{Analysis of Stochastic Geometry}\label{S3}
In this section, we use the Poisson point process (PPP) \cite{ref18} in SG theory to fit the behaviors of sub-UAVs in the A2A link. Besides, the analytic function is derived for the performance analysis. Taking the high-altitude airspace $V_h$ for example, all sub-UAVs follow the nearest neighbor association strategy\cite{ref19}, i.e.,

\begin{equation} 
  P\{d_{h}> R^\ast \}={\rm exp}(-λ\lambda_h V_h)={\rm exp}\bigg(-\frac{4}{3}\pi\lambda_h{d^3_{h}}\bigg),
\end{equation}

\noindent where $d_{h}$$\geq$0 and $R^\ast $ is the longest distance that the central UAV can serve. Therefore, the cumulative distribution function (CDF) of the distance $d_{h}$ from every sub-UAV $h$ in $\mathcal H$ to the central UAV $h_0$ is
\begin{equation} 
  F(d_{h})= P\{d_{h}\leq R^\ast \}=1-{\rm exp}\bigg(-\frac{4}{3}\pi\lambda_h{d^3_{h}}\bigg),
\end{equation}

\noindent and the probability density function (PDF) of $d_{h}$ is a derivative of CDF:
\begin{equation} 
  f(d_{h})=4\pi\lambda_h{d^2_{h}}{\rm exp}\bigg(-\frac{4}{3}\pi\lambda_h{d^3_h}\bigg).
\end{equation}

Supposing the transmission is succesful if the $\gamma_{h_i}$ is greater than the received threshold $\theta_h$  of the central UAV $h_0$, so the coverage probability of the $i$-$th$ sub-UAV in $\mathcal H$ is
\begin{equation} 
  P_{cov}^i=\mathbb{E}[P(\gamma_{h_i}\geq\theta_h|d_{h_i})].
  \label{eq19}
\end{equation} 
\noindent Since $\gamma_{h_i}$ is a function of $d_{h_i}$, $P_{cov}^i$ is further expressed as

\begin{equation}
        P_{cov}^i=\int_{0}^{\infty}P(\gamma_{h_i}\geq\theta_h|d_{h_i})f(d_h) \,d(d_h).
        \label{eq20}
\end{equation}

\par It is assumed that the average gain of small scale fading channel in the A2A is a random variable following the Gamma distribution with mean value of 1\cite{ref18}, which is depicted as

\begin{equation} 
  f(\rho)=\frac{\iota ^\iota    }{\Gamma(\iota )} \rho ^{\iota -1}e^{-\iota \rho }.
\end{equation}

\noindent When $\iota  $ is 1, the channel is considered as Rayleigh fading. $\rho $ follows an exponential distribution with mean value of 1. The PDF of $\rho $ is $f(x )$=$e^{-x}$, i.e., ${\rho _{h_i}}$$\sim$${\rm exp}(1)$ and ${\rho _{l_j}}$$\sim$${\rm exp}(1)$. Consequently, $P(\gamma_{h_i}\geq\theta_h|d_{h_i})$ in Eq. (\ref{eq19}) is further deduced as 
\vspace{-0.1cm} 
\begin{align}
  P(\gamma_{h_i}\geq\theta_h|d_{h_i})&=P\bigg({\rho_{h_i}}\geq\frac{\theta_h{{d^\delta _{h_i}}}(N+{P_s}S_{\mathcal{H}\backslash \{{h_i}\}}) }{P_sG_a}\bigg)\nonumber\\
  &={\rm exp}\bigg(\frac{-\theta_h{{d^\delta_{h_i}}}(N+{P_s}S_{\mathcal{H}\backslash \{{h_i}\}})}{P_sG_a}\bigg)\nonumber\\
  &={\rm exp}\bigg(\frac{-\theta_h{{d^\delta _{h_i}}}N}{P_sG_a}\bigg)\mathbb{L}_{S_{\mathcal{H}\backslash \{{h_i}\}}}\bigg(\frac{\theta_h{{d^ \delta _{h_i}}}}{G_a}\bigg).
  \label{eq22}
\end{align}
\noindent Let $\frac{\theta_h{{d^\delta_{h_i}}}}{G_a}=\Lambda_{h_i} $, and we have
\begin{equation}
  \mathbb{L}_{S_{\mathcal{H}\backslash \{{h_i}\}}}\bigg(\frac{\theta_h{{d^\delta_{h_i}}}}{G_a}\bigg)=\mathbb{L}_{S_{\mathcal{H}\backslash\{{h_i}\}}}(\Lambda_{h_i} )=\mathbb{E}[e^{-\Lambda_{h_i}(S_{\mathcal{H}\backslash \{{h_i}\}})}],
  \label{eq23}
\end{equation}
\noindent which is the Laplace transform of $S_{\mathcal{H}\backslash \{{h_i}\}}$, and is further derived as
\begin{align}
        \mathbb{L}_{S_{\mathcal{H}\backslash\{{h_i}\}}}&(\Lambda_{h_i})=\mathbb{E}\bigg[{\rm exp}\bigg(-\Lambda_{h_i}\sum_{{h}\in \mathcal{H}\backslash \{{h_i}\}}{G_a}{\rho _{h}}{d_{h}^{-\delta }}\bigg)\bigg]\nonumber\\
        &\overset{(\textbf{a})}{=}\mathbb{E}\bigg[\prod_{h\in \mathcal{H}\backslash\{{h_i}\}}\frac{1}{1+\Lambda_{h_i} G_ad_{h}^{-\delta }}\bigg]\nonumber\\
        &\overset{(\textbf{b})}{=}{\rm exp}\bigg[-\lambda_h\int_{V_h}\bigg(1-\frac{1}{1+\Lambda_{h_i} G_ad_{h}^{-\delta }}\bigg)d(d_{h})\bigg]\nonumber\\
        &\overset{(\textbf{c})}{=}{\rm exp}\bigg[-\lambda_h\int_{-L_x}^{L_x}\int_{-L_y}^{L_y}\int_{0}^{L_z}1-\nonumber\\
        &\quad\quad\frac{1}{1+\Lambda_{h_i} G_ad_{h}^{-\delta }}dxdydz\bigg]\nonumber\\
        &\overset{(\textbf{d})}{=}{\rm exp}(-\lambda_h\Theta_{h_i} ).
        \label{eq24}
      \end{align}
\noindent Wherein, (\textbf{a}) is obtained by the moment generating function, (\textbf{b}) follows the probability generating function (PGFL) of PPP\cite{ref20}, $d_{h}$ in (\textbf{c}) can be further expressed as $\sqrt{(x_{h}-x_{h_0})^2+(y_{h}-y_{h_0})^2+(z_{h}-z_{h_0})^2}$, and $\Theta_{h_i}$ in (\textbf{d}) represents the triple integral of (\textbf{c}).
By substituting (\ref{eq22}), (\ref{eq23}) and (\ref{eq24}) into (\ref{eq20}), $P_{cov}^i$ is calculated as 
\begin{align}
    P_{cov}^i&=\int_{0}^{\infty}4\pi\lambda_h{d^2_{h}}{\rm exp}\bigg(\frac{-\theta_h{{d^\delta_{h_i}}}N}{P_sG_a}\nonumber\\
    &-\lambda_h\Theta_{h_i}-\frac{4}{3}\pi\lambda_h{d^3_{h}}\bigg)d(d_{h}). 
\end{align}
The derivation process of formulas aforementioned is also applicable to the calculation of coverage probability for sub-UAVs in low-altitude airspace $V_l$.
\begin{algorithm}[t]
        \caption{On-board processing}
        \begin{algorithmic}[1]
                \renewcommand{\algorithmicrequire}{\textbf{Input:}}
                \renewcommand{\algorithmicensure}{\textbf{Output:}}
                \REQUIRE{$N^\ast $, $I_{h_i}^k$, $I_{h_i}^{k-1}$, $I_{h_i}^{k-2}$, $I_{h_i}^{k-3}$,..., $I_{h_i}^{k-N^\ast +1}$. }
                \ENSURE{$x_o$, $y_o$, and $z_o$.}
                \STATE $\mathcal M_{h_i}\gets\{0\}$.
                \STATE $n\gets1$.
                \WHILE {$ n\leq N^\ast $}
                  \STATE Update $m_{h_i}^n$ based on Eq. (\ref{eq27}) and (\ref{eq30}).
                  \STATE Update $\mathcal M_{h_i}$ based on Eq. (\ref{eq29}).
                  \STATE $n\gets n+1$.
                \ENDWHILE
                \STATE Calculate $m_{h_i}^{k-1}$ by putting $I_{h_i}^k$ and $I_{h_i}^{k-1}$ into Eq. (\ref{eq27}) and (\ref{eq30}).
                \IF{$m_{h_i}^{k-1} < \min \{m_{h_i}^1,m_{h_i}^2,...,m_{h_i}^n,...,m_{h_i}^{N^\ast }\}$}
                \STATE Abandon $I_{h_i}^k$.
                \STATE $\max \{m_{h_i}^1,m_{h_i}^2,...,m_{h_i}^n,...,m_{h_i}^{N^\ast }\} \gets m_{h_i}^{k-1}$.
        
                \ENDIF
                \IF{$m_{h_i}^{k-1} \geq  \max \{m_{h_i}^1,m_{h_i}^2,...,m_{h_i}^n,...,m_{h_i}^{N^\ast }\}$}
                \STATE $\min \{m_{h_i}^1,m_{h_i}^2,...,m_{h_i}^n,...,m_{h_i}^{N^\ast }\} \gets m_{h_i}^{k-1}$.
                \ENDIF
            \end{algorithmic}
            \label{A1}
 \end{algorithm}

\section{On-board processing based on MEC}\label{S4}
During the real transmission of flight information, there exist two abnormal conditions:
\begin{itemize}
        \item Redundant flight packet: Some packets of redundant information, which wastes the computing resources.
        \item Discrete flight packet: Packets loss caused by packets error or packets collision, which cause the discontinuity of the trajectory.
\end{itemize}
\par We propose an on-board processing mechanism to deal with the above two situations, i.e., the packet abandonment and packet supplement. We respectively leverage $Lon$, $Lat$ and $Alt$ to symbolize the longitude, latitude and altitude in a position packet. Hence, the position vector of the $i$-$th$ high-altitude UAVs in the $k$-$th$ ADS-B packets is classified as
\begin{equation}
        I_{h_i}^k=[Lon_{h_i}^k, Lat_{h_i}^k, Alt_{h_i}^k].
        \label{eq27}
\end{equation}
The following mechanisms can be applied to the high-altitude UAVs as well as the low-altitude UAVs.

\subsubsection{Mechanism of Abandonment}
The central UAV records the first $N^\ast $ position packets of the sub-UAV and compute the Minkowski distances of adjacent vectors in turns\cite{ref21}. The set of Minkowski distances $\mathcal M_{h_i}$ is defined as
\begin{equation}
        \mathcal M_{h_i}=\{m_{h_i}^1,m_{h_i}^2,...,m_{h_i}^n,...,m_{h_i}^{N^*}\}.
        \label{eq29} 
\end{equation}
The $n$-$th$ Minkowski distance $m_{h_i}^n$ in $\mathcal M_{h_i}$ is obtained as
\begin{align}
        m_{h_i}^n&=
        (|Lon_{h_i}^{n+1}-Lon_{h_i}^n|^p+|Lat_{h_i}^{n+1}-Lat_{h_i}^n|^p\nonumber\\
        &+|Alt_{h_i}^{n+1}-Alt_{h_i}^n|^p)^{1/p},
        \label{eq30}
\end{align}
where $p$ is the order of the Minkowski distances.
\par After receiving the $k$-$th$ packet from $h_i$, the central UAV calculates the Minkowski distance $m_{h_i}^{k-1}$ between $I_{h_i}^k$ and $I_{h_i}^{k-1}$. If
\begin{equation}
        m_{h_i}^{k-1}<\min \{m_{h_i}^1,m_{h_i}^2,...,m_{h_i}^n,...,m_{h_i}^{N^\ast }\},
\end{equation} 
the central UAV records $m_{h_i}^{k-1}$ in $\mathcal M_{h_i}$, and replaces the $\max \{m_{h_i}^1,m_{h_i}^2,...,m_{h_i}^n,...,m_{h_i}^{N^*}\}$, and abandons the $k$-$th$ packet.
\subsubsection{Mechanism of Supplement}
If 
\begin{equation}
m_{h_i}^{k-1}\geq\max \{m_{h_i}^1,m_{h_i}^2,...,m_{h_i}^n,...,m_{h_i}^{N^\ast }\},
\end{equation}
 the central UAV records $m_{h_i}^{k-1}$ in $\mathcal M_{h_i}$, and substitutes the $\min \{m_{h_i}^1,m_{h_i}^2,...,m_{h_i}^n,...,m_{h_i}^{N^\ast }\}$, and generate the packet supplement. Besides, the weighted average is used for the supplement. In detail, the on-board processing mechanisms are further described in Algorithm 1.

\section{Simulation Results and analysis}\label{S5}
We employ MATLAB to evaluate the detailed performance. The working frequency of the 5G system and ADS-B system is set as 3.5GHz and 1090MHz, respectively\cite{ref22}. Moreover, the bandwidth $B_{5G}$ and $B_{ADS-B}$ are set as 100MHz and 1MHz, respectively\cite{ref23}. Other detailed parameters in simulations are listed in TABLE \ref {tab1}.

\begin{figure}[t]
        \centerline{\includegraphics[width=1\linewidth]{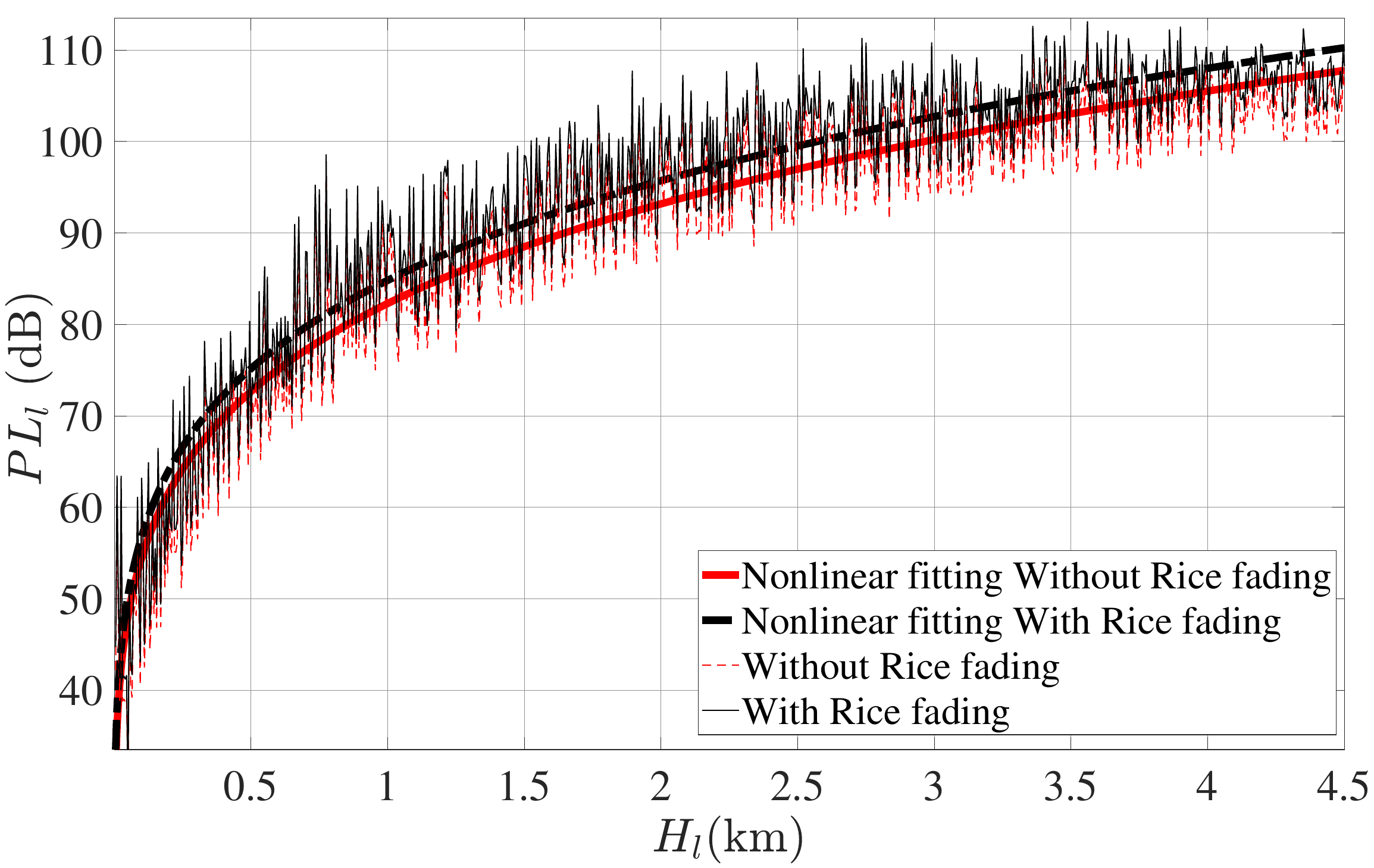}}
        \caption{Relationship between the $H_l$ and $PL_l$.}
        \label{f3}
\end{figure}
\begin{figure}[t]
        \centerline{\includegraphics[width=1\linewidth]{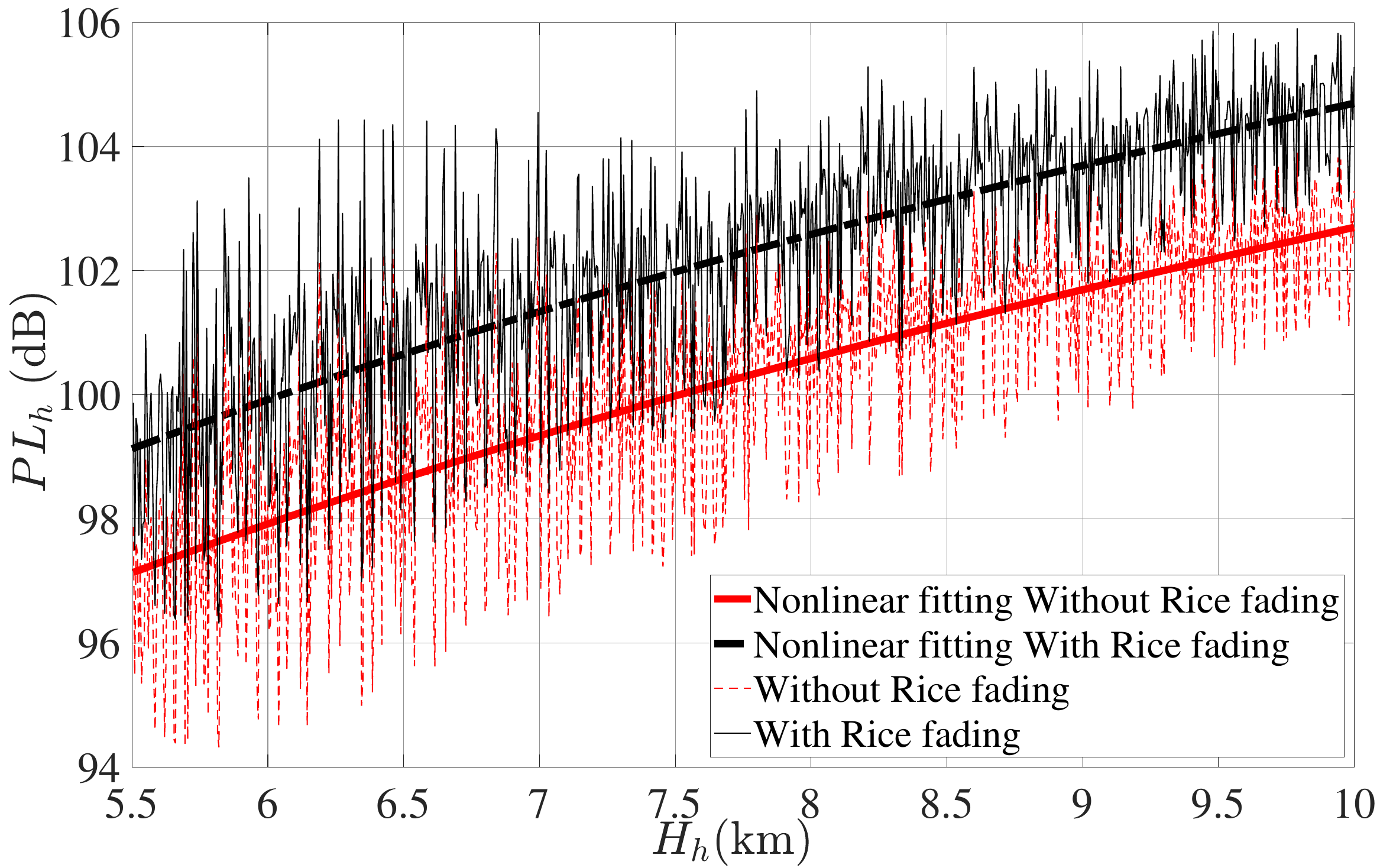}}
        \caption{Relationship between the $H_h$ and $PL_h$.}
        \label{f4}
\end{figure}
\begin{figure}[t]
        \centerline{\includegraphics[width=1\linewidth]{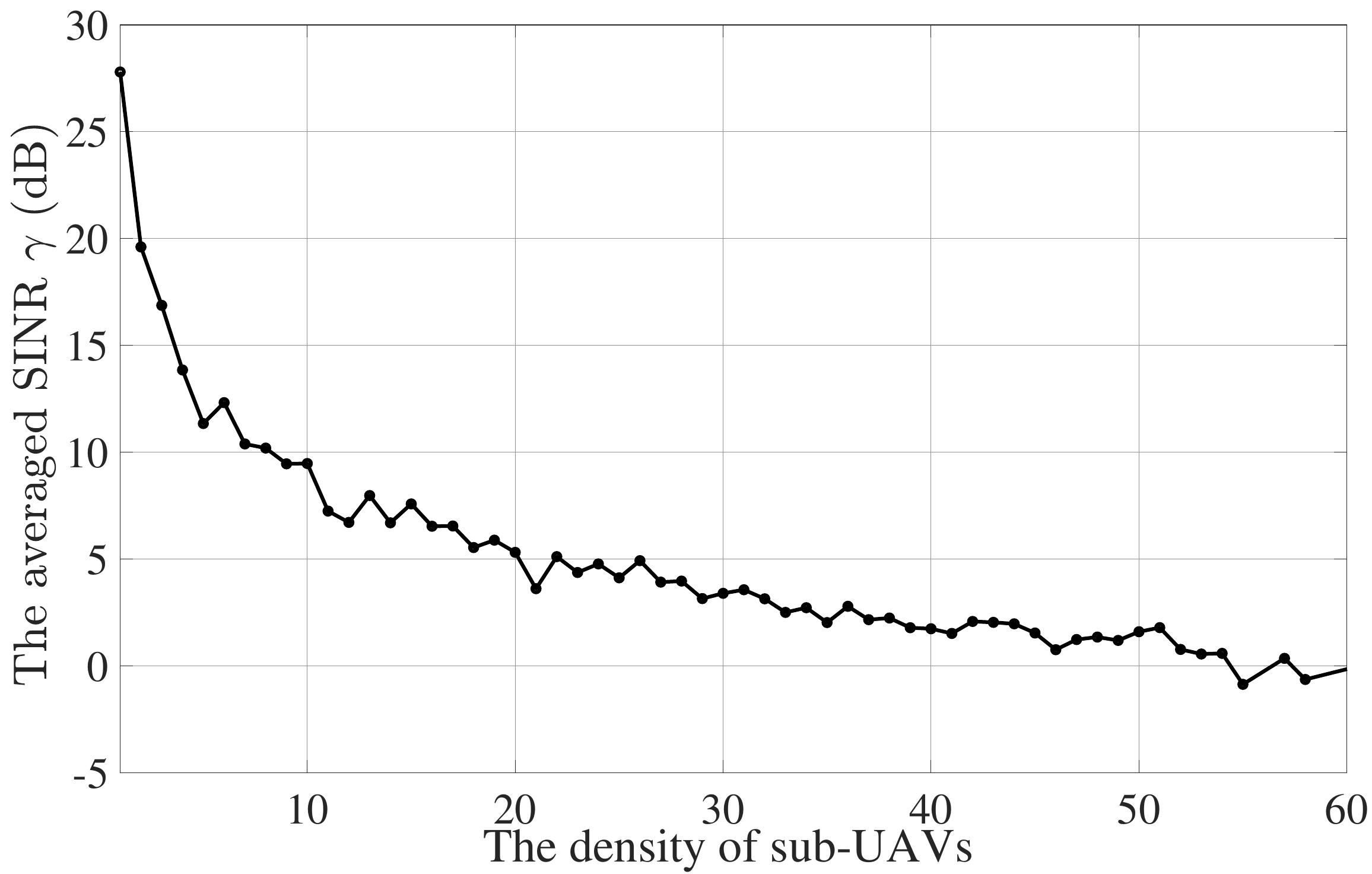}}
        \caption{Relationship between the density and averaged SINR.}
        \label{f5}
\end{figure}
\begin{figure}[t]
        \centerline{\includegraphics[width=1\linewidth]{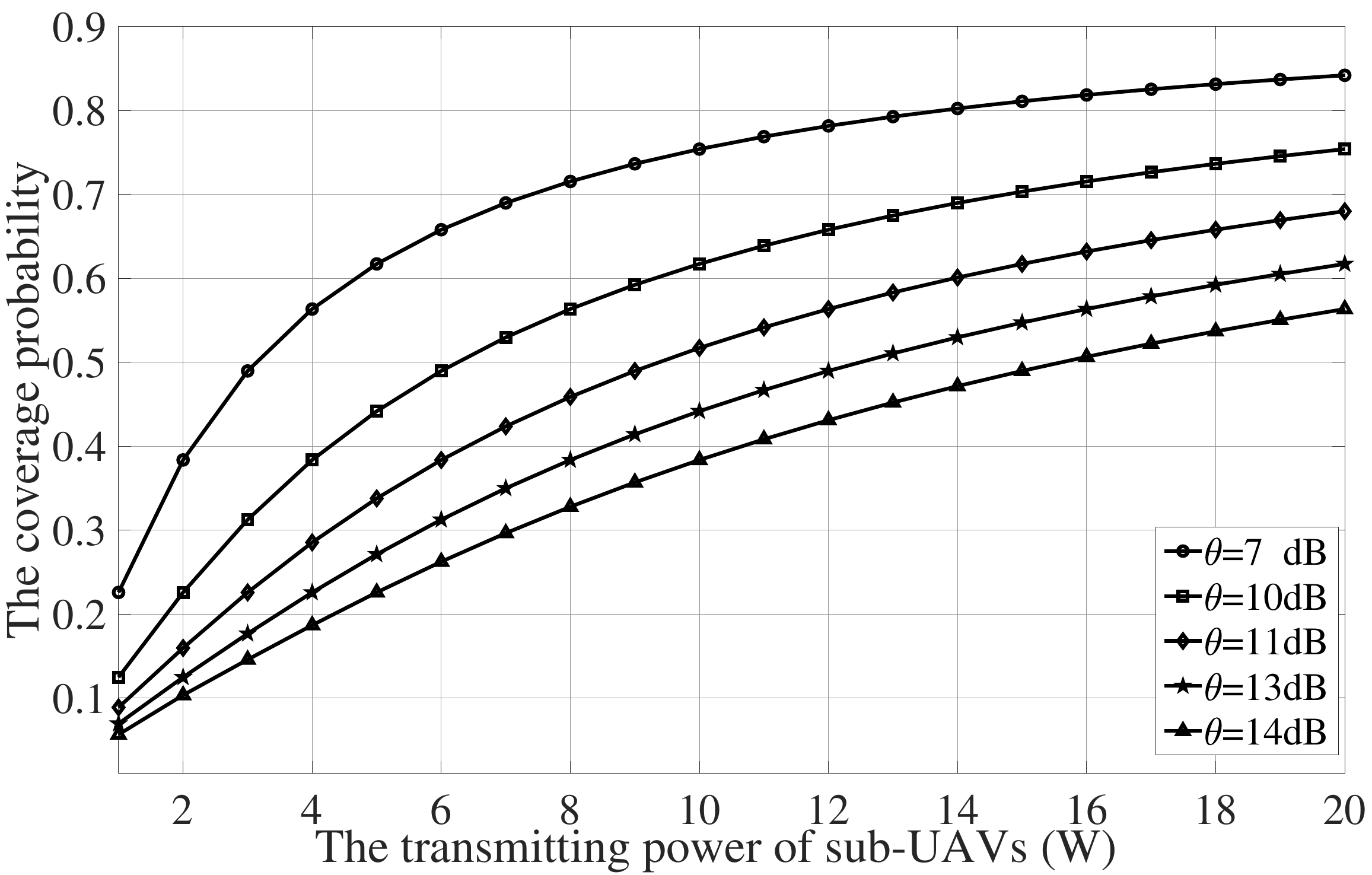}}
        \caption{Relationship between the transmitting power and coverage probability.}
        \label{f6}
\end{figure}

\begin{table}[t]
        \centering
        \caption{Key Parameters in the Simulations}
        \begin{center}
        \begin{tabular}{|p{2.8cm}<{\centering}|p{2.8cm}<{\centering}|}
        \hline  
        Parameter &Value\\
        \hline
        $f_{5G}$, $f_{ADS-B}$&3.5GHz, 1090MHz\\
        \hline
        $\lambda_{5G}$, $\lambda_{ADS-B}$& 0.0857m, 0.2752m \\
        \hline
        $B_{5G}$, $B_{ADS-B}$& 100MHz, 1MHz \\
        \hline
        $n_0$& -174dBm/Hz \\
        \hline
        $P_s$& 1W$\sim $20W\\
        \hline
        $P_{ch}$, $P_{cl}$& 20W, 20W \\
        \hline
        $G_g$, $G_a$& 20dBi, 23dBi\\
        \hline
        $\theta_h$, $\theta_l$& 7dB$\sim $14dB, 7dB$\sim $14dB\\
        \hline
        $\delta$& 2$\sim $4.9\\
        \hline
        $λ\lambda_l $, $λ\lambda_h $& 1$\sim $60, 1$\sim $60\\
        \hline
        $\Delta H$ & 4.5km\\
        \hline
        $H_0$, $H_G$ & 1km, 50m \\
        \hline
        $\sigma$ & 5$\times10^3$ \\
        \hline
        $\varepsilon_r$ & 15\\
        \hline
        \end{tabular}
        \label{tab1}
        \end{center}
\end{table}

\begin{figure}[t]
        \centerline{\includegraphics[width=1\linewidth]{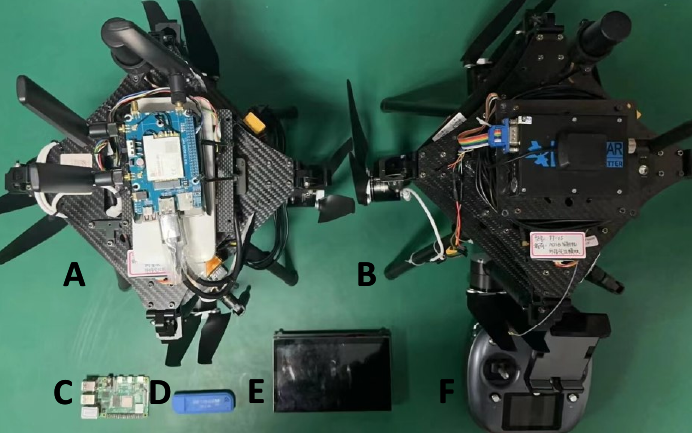}}
        \caption{Experimental equipments.}
        \label{f7}
\end{figure}

\begin{figure*}[t] 
        \centering
        \begin{minipage}{1\linewidth}
        \centering    
        \subfloat[Original data.]{
        \includegraphics[width=0.49\linewidth]{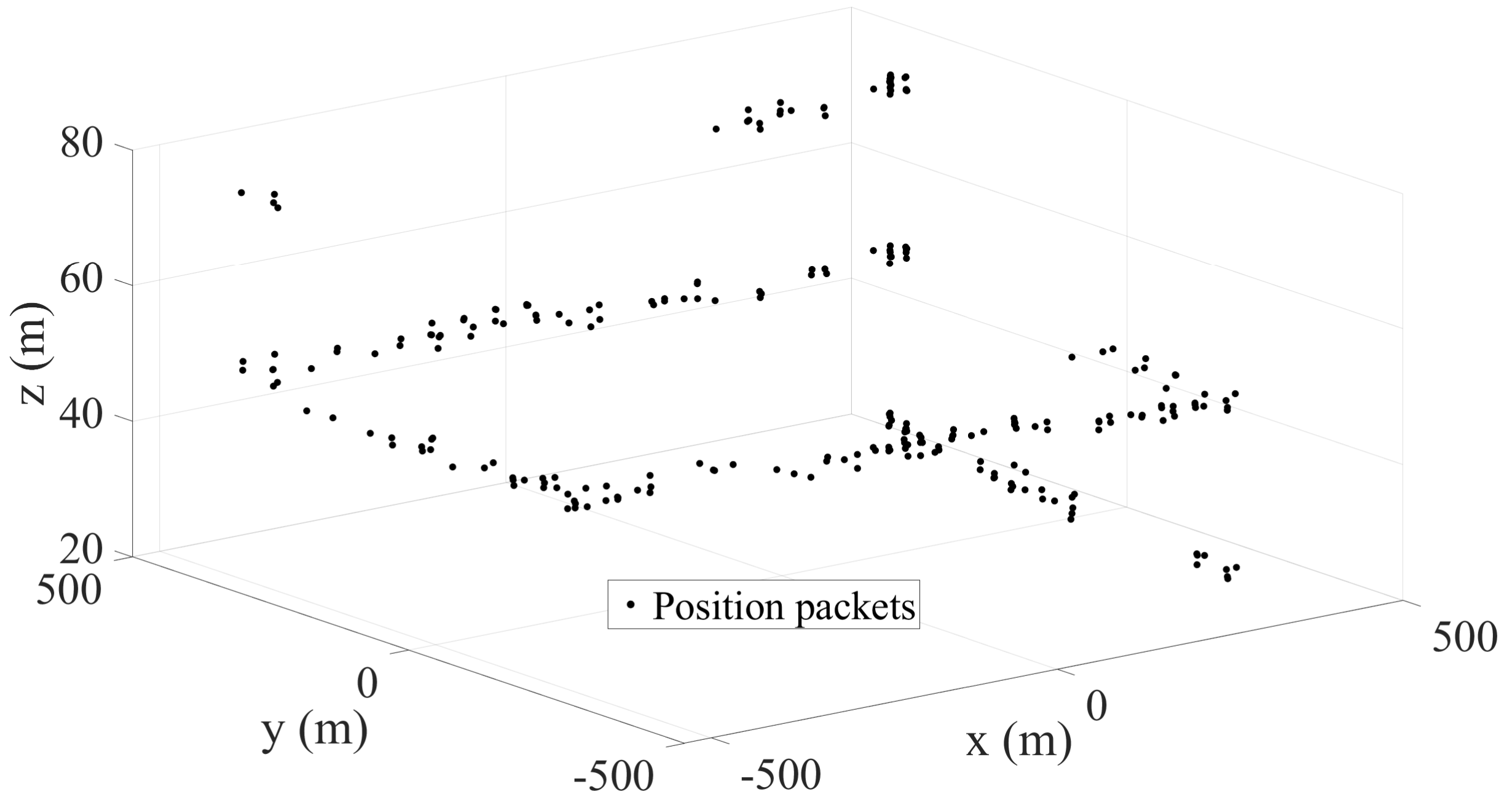}
        \label{f8.1}}
        \subfloat[Optimized data.]{
        \includegraphics[width=0.49\linewidth]{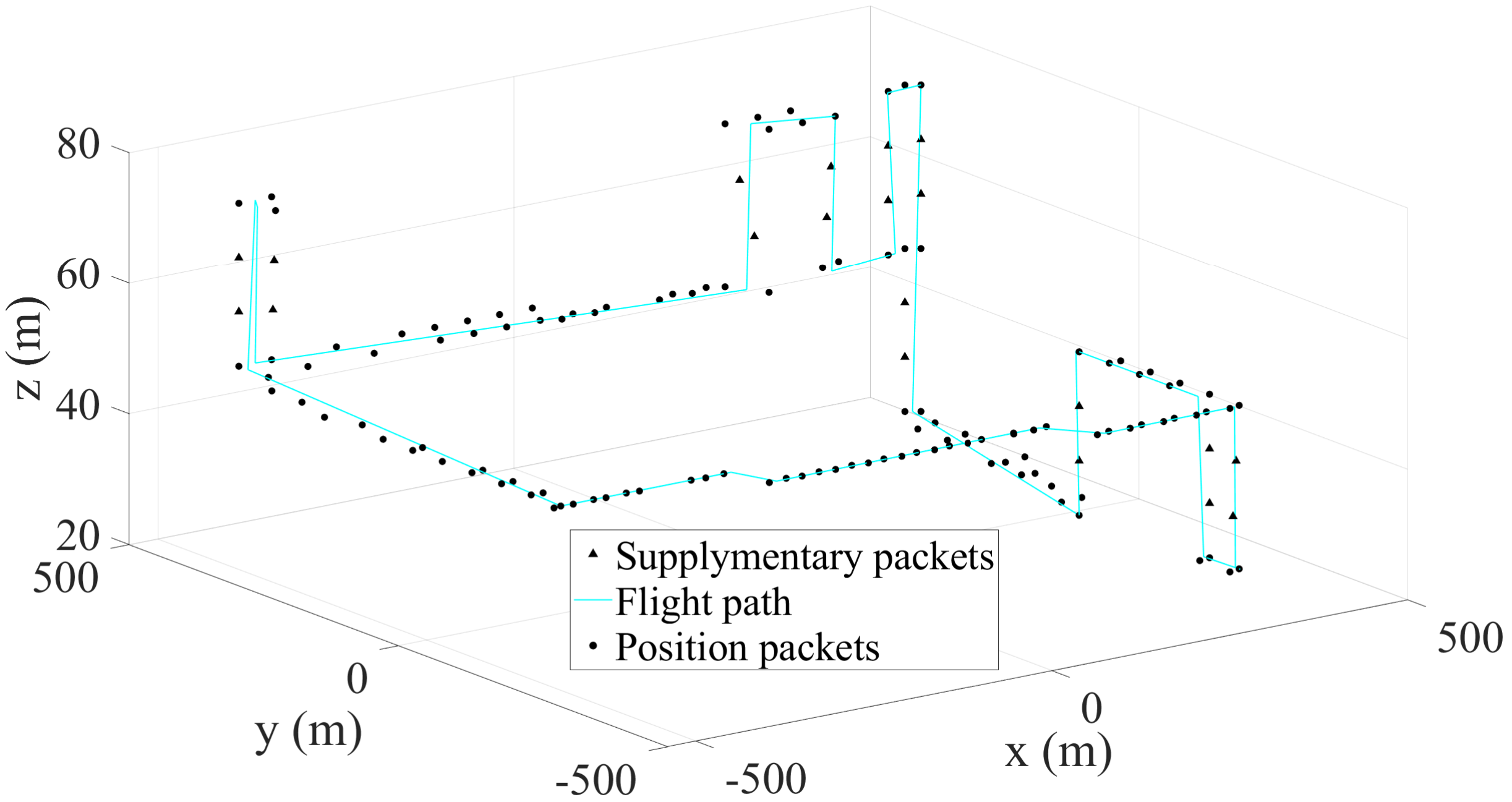}
        \label{f8.2}}
        \caption{On-board processing based on MEC.}
        \label{f8}
        \end{minipage}
\end{figure*}
\par Fig. \ref {f3} depicts the A2G channel performance for the low-altitude central UAV $l_0$. The black solid thin line indicates the path loss without Rice fading while red dotted thin line demonstrates the path loss with Rice fading. The thick lines are corresponding nonlinear regression curves. From Fig. \ref {f3}, the A2G models are oscillation models, following the law of free space path loss. With the height $H_l$ of the central UAV $l_0$ increasing, the path loss $PL_l$ also increases. After the introduction of Rice fading, $PL_l$ is growing, but the increment is not obvious in low-altitude airspace. When $H_l$ is greater than 4km, the $PL_l$ curve tends to be smooth. The height of the central UAV $l_0$ is recommended to be limited in respect to the practical requirement of path loss.

\par Fig. \ref {f4} illustrates the A2G channel performance for the high-altitude central UAV $h_0$. The black solid thin line represents the path loss without Rice fading while the red dotted thin line symbolizes the path loss with Rice fading. The thick lines are corresponding nonlinear regression curves. It is observed from Fig. \ref {f4}, $PL_h$ magnifies as $H_l$ enlarges, but the increment is not obvious in the high-altitude airspace in respect to the ADS-B central UAV $h_0$. Further, the difference in high-altitude airspace is obvious with Rice fading compared to without it. Since the height limitation of the central UAV $h_0$ is within $(5.5{\rm km}, 10{\rm km})$, the growth of $PL_h$ does not exhibit a smooth trend. The height of central UAV $h_0$ should be determined according to the actual requirement of GS.

\par Fig. \ref {f5} shows the relationship between the density of sub-UAVs and averaged SINR of the signal received by the central UAV. Both airspaces have the same size, and all sub-UAVs send packets to the central UAVs, so the performance analysis of the A2A does not distinguish between the high-altitude and the low-altitude. As the density of sub-UAVs increases, the averaged SINR decreases. The increment of sub-UAVs leads to abundant signals propagating in the airspace. Thus, there exist more interference signals when the central UAV receives a specific signal. When the density of sub-UAVs exceeds 30, the averaged SINR of the signals degenerates to around 0, which is disadvantageous to signal processing. Therefore, the density of sub-UAVs should adapt to local conditions. 
\par Fig. \ref {f6} describes the relationship between the coverage probability and transmitting power of sub-UAVs under different received thresholds. Moreover, different icons symbolize different received thresholds. The path loss and density are set as 2 and 20, respectively. With the power rising, the signal coverage probability grows. Besides, when the power is greater than 16W, it shows a smooth trend in the increment of the coverage probability regardless of thresholds. However, the enlargement of the threshold leads the coverage probability to fall. Supposing the threshold is 7dB, fixing the transmitting power as 8W to make the highest efficiency of energy investment, since the gain of the coverage probability from the increment of a unit transmitting power is continuously decreasing.

\par As shown in Fig. \ref {f7}, we collect the real flight data of the UAV for algorithm verification. \textbf{A} is the UAV equipped with 5G module (SIM8262E-M2), and \textbf{B} is the UAV equipped with ADS-B out device (PT050X). Additionally, \textbf{C} is the Raspberry PI (4-Model-B), playing the role of GS and processing the position packets. \textbf{D} is the Realtek software defined radio as the antenna to receive the position packets. Besides, \textbf{E} is the display and \textbf{F} is the UAV controller. The sending and receiving terminals are constructed in our lab. Further, in the experiment, we consider A and B as sub-UAVs and C as the central UAV.
\par Fig. \ref {f8.1} and Fig. \ref {f8.2} show the original data and optimized data of trajectory \ding{172}. The dots represent the received position packets, and the triangles represent the supplementary packets. As seen from Fig. \ref {f8.1}, there exist abundant redundant position packets. Besides, discontinuities of trajectory appears when the UAV changes the flight altitude. After the on-board processing mechanism, corresponding to Fig. \ref {f8.2}, the trajectory is more clear and smooth. In this scenario, the redundant data is effectively reduced by 52.55\%, and 20 supplementary packets are added. 

\vspace{-0.3cm}  

\section{Conclusions}\label{S6}
In this paper, we build the cooperating framework of ADS-B in 5G for hierarchical aerial networks. Then, we present the system model of the A2G and A2A channels based on deterministic modeling and stochastic modeling, respectively. Further, we analyze the performance of the A2G and A2A systems, and provide the corresponding suggestions. Finally, we propose the on-board processing mechanism based on MEC and verify the algorithm by simulations as well as experiments. The results indicate that the algorithm effectively filters out redundant packets, i.e., more than 50\% of redundant packets is filtered out. This paper helps to strengthen the flight safety of UAVs and enhance the competence of airspace flow management. In the future work, we will utilize the on-board processing mechanisms to verify various flight trajectories. Besides, additional attention will be concentrated on the energy consumption of different communication modes.

\vfill
\end{CJK}
\end{document}